\documentclass[twocolumn]{svjour3}          

\smartqed  

\usepackage{amsmath,amssymb}
\usepackage{graphicx}
\usepackage[all]{xy}
\usepackage{txfonts}
\usepackage{hyperref,color}
\usepackage{xcolor,colortbl}
\definecolor{LightCyan}{rgb}{0.88,1,1}

\numberwithin{equation}{section}

\newcommand{\R}{\mathbb{R}}

\renewcommand{\ell}{\mathrm{L}}

\usepackage[mathlines]{lineno}

\newcommand*\patchAmsMathEnvironmentForLineno[1]{%
  \expandafter\let\csname old#1\expandafter\endcsname\csname #1\endcsname
  \expandafter\let\csname oldend#1\expandafter\endcsname\csname end#1\endcsname
  \renewenvironment{#1}%
     {\linenomath\csname old#1\endcsname}%
     {\csname oldend#1\endcsname\endlinenomath}}%
\newcommand*\patchBothAmsMathEnvironmentsForLineno[1]{%
  \patchAmsMathEnvironmentForLineno{#1}%
  \patchAmsMathEnvironmentForLineno{#1*}}%
\AtBeginDocument{%
\patchBothAmsMathEnvironmentsForLineno{equation}%
\patchBothAmsMathEnvironmentsForLineno{align}%
\patchBothAmsMathEnvironmentsForLineno{flalign}%
\patchBothAmsMathEnvironmentsForLineno{alignat}%
\patchBothAmsMathEnvironmentsForLineno{gather}%
\patchBothAmsMathEnvironmentsForLineno{multline}%
}

\begin{document}

\title{Overview of image-to-image translation by use of deep neural networks: denoising, super-resolution, modality conversion, and reconstruction in medical imaging}
\titlerunning{Image-to-image translation in medical imaging}

\authorrunning{S. Kaji and S. Kida}
\author{Shizuo Kaji\thanks{The two authors contributed equally to this work.} \and Satoshi Kida$^\ast$}

\institute{
S. Kaji \at
              Institute of Mathematics for Industry, Kyushu University, Japan / JST PRESTO\\
              \email{skaji@imi.kyushu-u.ac.jp}
            \and
S. Kida \at
              LPixel Inc. / The University of Tokyo hospital \\
              \email{satoshikida1492@gmail.com}          
           }
\date{Received: date / Accepted: date}

\maketitle

%

\begin{abstract}

Since the advent of deep convolutional neural networks (DNNs),
computer vision has seen an extremely rapid progress 
that has led to huge advances in medical imaging.
Every year, many new methods are reported of in conferences such as
the International Conference on Medical Image Computing and Computer Assisted Intervention (MICCAI) and
Machine Learning for Medical Image Reconstruction (MLMIR),
or published online at the preprint server arXiv.
There is a plethora of surveys on applications of neural networks in medical imaging (see \cite{Sahiner2018DeepLI} for a relatively recent comprehensive survey).
This article does not aim to cover all aspects of the field but
 focuses on a particular topic, \emph{image-to-image translation}. 
Although the topic may not sound familiar, it turns out that 
many seemingly irrelevant applications can be understood as instances of image-to-image translation.
Such applications include (1) noise reduction, (2) super-resolution, (3) image synthesis, and (4) reconstruction.
The same underlying principles and algorithms work for various tasks.
Our aim is to introduce some of the key ideas on this topic from a uniform point of view. 
We introduce core ideas and jargon that are specific to image processing by use of DNNs.
Having an intuitive grasp of the core ideas of and a knowledge of technical terms
would be of great help to the reader for understanding the existing and  future applications.

Most of the recent applications which build on image-to-image translation are based on one of two fundamental architectures, called pix2pix and CycleGAN, depending on 
whether the available training data are \emph{paired} or \emph{unpaired}
(see \S \ref{sec:AEN}).
We provide codes (\cite{github1,github2}) which implement these two architectures with various enhancements.
Our codes are available online with use of the very permissive MIT license.
We provide a hands-on tutorial for training a model for denoising based on our codes (see \S \ref{sec:sample}).
We hope that this article, together with the codes, will provide both an overview and the details of the key algorithms, and that it will serve as a basis for the development of new applications.


\end{abstract}

\section{Image processing with deep neural networks}
We begin by considering basic notions about neural networks in general.
The reader may want to skip to \S \ref{sec:AEN}, where
we discuss our main topic of image-to-image translation.

Conventional medical image processing algorithms rely on domain specific knowledge and assume some underlying model, and hence,
each algorithm is tailored for a specific task.
On the other hand, in recent years purely data-driven approaches without specific models have become popular in medical imaging 
(e.g., \cite{DBLP:series/acvpr/978-3-319-42998-4,DBLP:conf/miccai/2018mlmir,Litjens2017ASO}).
In particular, \emph{deep neural networks (DNNs)}  have proved to be very powerful in various image processing tasks.
The most important fact for image processing is that
an image is expressed by a real-valued vector.
A greyscale image of dimension $w\times h$ is an element of $\R^{w\times h}$,
where $\R^{w\times h}$ is the set of real-valued vectors of dimension $w\times h$.
Similarly, a full color picture of dimension $w\times h$ is an element of $\R^{w\times h\times 3}$
consisting of three channels corresponding to Red, Green, and Blue.
This fact allows us to apply various mathematical operations to the processing of images.
We can apply a mapping $\R^{w_1\times h_1}\to \R^{w_2\times h_2}$ to translate one greyscale image to another.
Image filters such as Sobel filters, FFT, and bilateral filters are handcrafted mappings\footnote{Various image filters are implemented in the free software Fiji \cite{fiji}, and we can easily try them out to see their characteristics.},
whereas DNNs are meant for finding of useful filters automatically from a large amount of data.
DNN-based methods have achieved state-of-the-art performance in many image translation tasks, including denoising, super-resolution, image synthesis, and reconstruction.

\subsection{Regression with neural networks}
In this subsection, we explain what are neural networks and deep learning.
A good introduction to deep learning in general is given in \cite{nielsenneural}.
For those who are interested in the theory of deep learning in depth, we refer to \cite{GoodBengCour16}.

Neural networks are just a particular class of multi-variate functions $f_w: \R^n \to \R^m$ with parameters (called \emph{weights})
$w\in \R^l$.
The parameters $w$ are \emph{learnable}, so that they are determined by the given data through \emph{training}
for the solution of a given problem.
The simplest example is the linear function $f_{(w_1,w_0)}: \R^1 \to \R^1$, defined by 
$f_{(w_1,w_0)}(x)=w_1x+w_0$, whose weights are $w=(w_1,w_0)\in \R^2$.

A neural network can be used to approximate (\emph{fit} or \emph{regress}) an unknown function $g$ from a finite number of its input-output pairs,
\begin{equation*}
  \{ (x_1,g(x_1)), (x_1,g(x_1)), \ldots, (x_k,g(x_k))\}.
\end{equation*}
Here, $k$ is the number of input-output pairs which are observed and known to us.
For a new $x$ different from $x_1,\ldots,x_k$,
we want to use $f_w(x)$ as a prediction of the unknown value $g(x)$.

To measure how well $f_w$ approximates $g$, we need a \emph{loss} function. A popular choice is the squared
\emph{$\ell^2$-distance} $L(f_w,g)=\sum_{i=1}^k |f_w(x_i)-g(x_i)|^2$. 
A loss function is designed so that,
the smaller $L(f_w,g)$ is, the better $f_w$ approximates $g$ with respect to the loss function $L$.
Hence, we try to find parameters $w$ which minimize $L(f_w,g)$, or at least, 
which make  $L(f_w,g)$ reasonably small.
In the case of the simplest neural network $f_{w_1,w_0}(x)=w_1x+w_0$
with the squared $\ell^2$-distance as the loss function, this is achieved by  ordinary least-squares fitting.
For expressing functions more complex than simple linear functions, 
the crucial idea of neural networks is to stack (or compose) linear functions.
Of course, the composition of two linear functions is again a linear function.
So we insert the \emph{activation function} (or the \emph{non-linearity})
between linear functions.
The simplest two-layer neural network $\R^1\to \R^1$ is given by
\begin{linenomath}\begin{equation*}
f_{w}(x)= w_{2,1}\max(w_{1,1}x+w_{1,0},0) + w_{2,0}.
\end{equation*}\end{linenomath}
This looks complicated, but it is just the composition of two linear functions sandwiching 
the \emph{ReLU} (Rectified Linear Unit) activation, which outputs
the input itself if it is positive and outputs zero otherwise:
\begin{linenomath}\begin{align*}
\R^1 & \to & \R^1 \qquad & \xrightarrow{\mathrm{ReLU}}& \R^1 \qquad \\
x &\mapsto& y_1=w_{1,1}x+w_{1,0} &\mapsto& y_2=\max(y_1,0)  \\
& & &\to& \R^1 \qquad \\
& & &\mapsto& y_3=w_{2,1}y_2+w_{2,0}.
\end{align*}\end{linenomath}
The point is that composition increases the expressive power drastically
(see, for example, \cite{Safran2017}),
while keeping the building blocks rather simple (consisting of only linear functions and activation functions).
Neural networks with three or more layers are referred to as \emph{deep} neural networks (DNNs).
The \emph{universal approximation theorem} (e.g., \cite{universal}) roughly states that almost any 
function $g: \R^n \to \R^m$ which appears in practical applications can be approximated arbitrarily well by a deep neural network if we allow the number of weights to be
arbitrarily large.
Nowadays, the number of layers of a DNN is of the order of dozens and more, and 
the parameter $w$ exists in a high dimensional space $\R^k$ with $k>10,000,000$.
\begin{remark}
If three-layer networks suffice for universal approximation, why do we need 
dozens of layers?
In practice, deeper neural networks are better at various jobs than are 
shallower and wider neural networks with the same number of weights.
However, in theory, deep narrow neural networks in which the width is bounded are not universal\cite{NIPS2017_7203}.
This fact poses an interesting question in how to design the structure of 
neural networks for a particular task.
\end{remark}

\subsection{Convolutional neural networks}
A linear map (without a constant term) $\R^n \to \R^m$ 
is specified by an $m\times n$ matrix\footnote{
Usually, a linear layer involves the constant term as well, so that 
it has the form $x \mapsto Ax + b$ for $b\in \R^m$. The term $b$ is often referred to as the \emph{bias}.}.
This means that the number of weights is $mn$, which is huge;
e.g., when $n=m=256 \times 256$, the figure is over 4 million.
In principle, the larger the number of weights, the more data we need to find good values for them by training.
Thus, we have to keep the number of weights relatively small based on some prior knowledge about images in general.
It is reasonable to assume that two pixels far apart are independent,
but pixels that are next to each other are highly correlated.
Also, the correlation does not depend on the absolute location of the pixel in a picture, but
should be similar everywhere.
A mathematical tool called the \emph{convolution} is what we need to exploit this observation.
There are many good expository articles on 2D \emph{convolutional neural networks} (CNNs) (e.g., \cite{dumoulin2016guide}).
Here, we present an illustrative toy example of 1D convolution.
Basically, we can think of convolution as a filter application with a sliding window.
Assume that we have an input vector $x=(x_1,x_2,\ldots,x_n)\in \R^n$.
Given another vector $w=(w_1,w_2,w_3)$ called the \emph{kernel},
we compute
\begin{align}\label{eq:convolution}
 w*x = & (w_1x_1+w_2x_2+w_3x_3, w_1x_2+w_2x_3+w_3x_4, \ldots, \nonumber\\
 & w_{1}x_{n-2}+w_{2}x_{n-1}+w_3x_n) \in \R^{n-2}.
 \end{align}
This mapping $x\mapsto w*x$ is called the convolution with kernel $w$, and it defines a linear mapping
$\R^n \to \R^{n-2}$ with weight $w$. 
Note that the number of weights is fixed to three and is independent of the dimension of the input $x$.
The trick is that the weights are
\begin{itemize}
\item \emph{sparse}; each entry of the output $w*x$ depends only on three entries, $x_{i-1},x_i,x_{i+1}$, in the input,
\item \emph{shared}; the same weights, $w_1,w_2,w_3$, are used for all entries of the output.
\end{itemize}

Next, we explain some more technical aspects of convolution.
First, if we want to keep the input and the output vector sizes the same, 
we can use \emph{padding} $(x_1,x_2,\ldots,x_n) \mapsto (0,x_1,x_2,\ldots,x_n,0)$,
so that the output \eqref{eq:convolution} becomes
\begin{align*}
 (w_2x_1+w_3x_2, w_1x_1+w_2x_2+w_3x_3, w_1x_2+w_2x_3+w_3x_4, \ldots,\\ 
 w_{1}x_{n-2}+w_{2}x_{n-1}+w_3x_n, w_{1}x_{n-1}+w_{2}x_{n}) \in \R^{n}.
\end{align*}
Padding also prevents the boundary elements from being treated lightly; 
without padding, $x_1$ and $x_n$ contribute only once to the output, whereas $x_3,\ldots,x_{n-2}$ appear three times.

A layer with the general form of a linear map $\R^n\to \R^m$
is said to be \emph{fully connected}, in contrast to being convolutional.
We can stack convolutional and/or fully connected layers by sandwiching activation functions to form a DNN.
With a fully connected layer, we can choose an arbitrary output dimension $m$.
We have some but limited control over the size of the output of a convolutional layer.
If we want to down-sample the signal, 
we use convolution with a \emph{stride}.
For example, with stride 2 and a padding, the output \eqref{eq:convolution} becomes
\begin{align*}
 (w_2x_1+w_3x_2, w_1x_2+w_2x_3+w_3x_4, w_1x_4+w_2x_5+w_3x_6, \ldots,\\ 
 w_{1}x_{n-3}+w_{2}x_{n-2}+w_3x_{n-1}, w_{1}x_{n-1}+w_{2}x_{n}) \in \R^{ (n+1)/2 }
\end{align*}
when $n$ is odd and
\begin{align*}
 (w_2x_1+w_3x_2, w_1x_2+w_2x_3+w_3x_4, w_1x_4+w_2x_5+w_3x_6, \ldots,\\ 
 w_{1}x_{n-4}+w_{2}x_{n-3}+w_3x_{n-2}, w_{1}x_{n-2}+w_{2}x_{n-1}+w_3x_{n}) \in \R^{ n/2 }
\end{align*}
when $n$ is even.
Down-sampling can also be achieved by a fixed (not learnable) filter such as  \emph{max-pooling} and  \emph{average-pooling} with a stride.

We can also deal with multi-channel signals.
For example, for a two-channel 1D signal
$x=\begin{pmatrix} x_{11},x_{12},\ldots,x_{1n} \\ x_{21},x_{22},\ldots,x_{2n} \end{pmatrix}\in \R^{2 \times n}$,
we use a kernel 
$w=\begin{pmatrix} w_{11},w_{12},x_{13} \\ w_{21},w_{22},w_{23} \end{pmatrix}$,
and the convolution is defined to be
\begin{align*}
w*x  =( &
(w_{11}x_{11}+w_{12}x_{12}+w_{13}x_{13})+(w_{21}x_{21}+w_{22}x_{22}+w_{23}x_{23}), \\
& (w_{11}x_{12}+w_{12}x_{13}+w_{13}x_{14})+(w_{21}x_{22}+w_{22}x_{23}+w_{23}x_{24}), \\
& \ldots, (w_{11}x_{1(n-2)}+w_{12}x_{1(n-1)}+w_{13}x_{1n}) \\ & +(w_{21}x_{2(n-2)}+w_{22}x_{2(n-1)}+w_{23}x_{2n})
)
 \in \R^{n-2}.
\end{align*}
We can also use multiple kernels to make an output multi-channel;
the number of output channels is equal to the number of kernels used.
For example, multiple gradient-like filters in different orientations may be useful for capturing image features.
Each channel in the output is sometimes called the \emph{feature map}.
With CNNs, kernels (filters) are not designed by a human being, but are learned from data.
To \emph{train} a CNN, we have to provide its objective in terms of 
a \emph{loss function} (or a cost function, an energy function, a penalty function).
Training means \emph{optimizing} the loss function 
by finding a set of weights which attains a small value of the given loss function.
The training proceeds iteratively by gradual adjustment of the weights to lower the value of the loss functions. This usually requires much time and powerful equipment including GPUs.

\begin{remark}
An 1D signal with multiple channels should not be confused with a
2D signal with a single channel.
Convolution operates differently. Namely, kernels slide only spatially, but not in the direction of channels.
\end{remark}

Usually, a down-sampling layer comes with more output channels than input channels.
The numbers are chosen so that the overall size of the output is smaller than 
that of the input, and the layer serves as a compressor of information.
Learning a compact representation in this manner is one of the key ideas of CNN.
For 2D signals, a typical down-sampling layer has stride two with doubling of output channels,
so that the total size of information is halved.

If we want to up-sample the signal, 
we can for example, use simple bilinear interpolation
or rearrange multiple channels \cite{pixelshuffler}.
It should be noted that it has been popular to use
\emph{deconvolution} (or \emph{transposed convolution}) for up-sampling,
which is the linear adjoint to convolution (see \cite{dumoulin2016guide} for details),
However, deconvolution often results in a draughtboard-like noise in the output
\cite{deconvnoise}.

Often, between a convolution and an activation, various \emph{normalization} methods are
 applied for better convergence in optimization \cite{batchnorm}. 
In addition, in image translation, normalization is known to have a great impact on the output \cite{instancenorm}.
The popular combination of convolution followed by batch normalization followed by ReLU activation is abbreviated as CBR.

In image processing, the majority of neural networks are \emph{feed-forward networks}, so that the data flow in one direction and there is no feedback loop.
Designing a DNN architecture mainly involves knowing how to connect layers 
with different numbers of channels, strides, and kernel sizes.
A particular attention should be paid to the \emph{receptive field}
when designing an image-to-image translation network;
each pixel in the output image is affected by not all pixels in the input image but by a patch in it. The size of the receptive field can be easily computed by a dedicated calculator which is found online.

In summary, employing DNNs for a specific task requires
\begin{enumerate}
\item collecting training data (this is probably the hardest part)
\item designing the network architecture of DNNs, and loss functions (there is a variety of off-the-shelf architecture that one can choose)
\item training (time-consuming, powerful hardware is often required).
\end{enumerate}
Once a trained (learned) model, which consists of DNNs with trained weights,
is obtained, using it is not very demanding. 
We only need the model and do not need the training data.
The process of getting outputs from a trained model is called \emph{inference}, and 
it usually does not require a great deal of machine power.
However, it has to be noted that some information on
the training data can be extracted from a trained model
by a malicious attacker \cite{Fredrikson2015}.
Therefore, care has to be taken when one provides a learned model which is trained on data that contain sensitive personal information.

\subsection{Image translation with encoder-decoder networks and GANs}
\label{sec:AEN}
Now we focus on the way to apply DNNs for image-to-image translation.
Two powerful inventions, \emph{encoder-decoder networks} (e.g, \cite{unet})
and \emph{generative adversarial networks (GANs)} \cite{GAN},
have accelerated the use of DNNs in image-to-image translation (\cite{ganlist}). 
A celebrated paper \cite{Isola2016} crystallizes how these are combined to yield image-to-image translation,
and we highly recommend that the reader have a look at it.
We give an informal account of these two key inventions.

\begin{figure*}
  \includegraphics[width=0.94\textwidth]{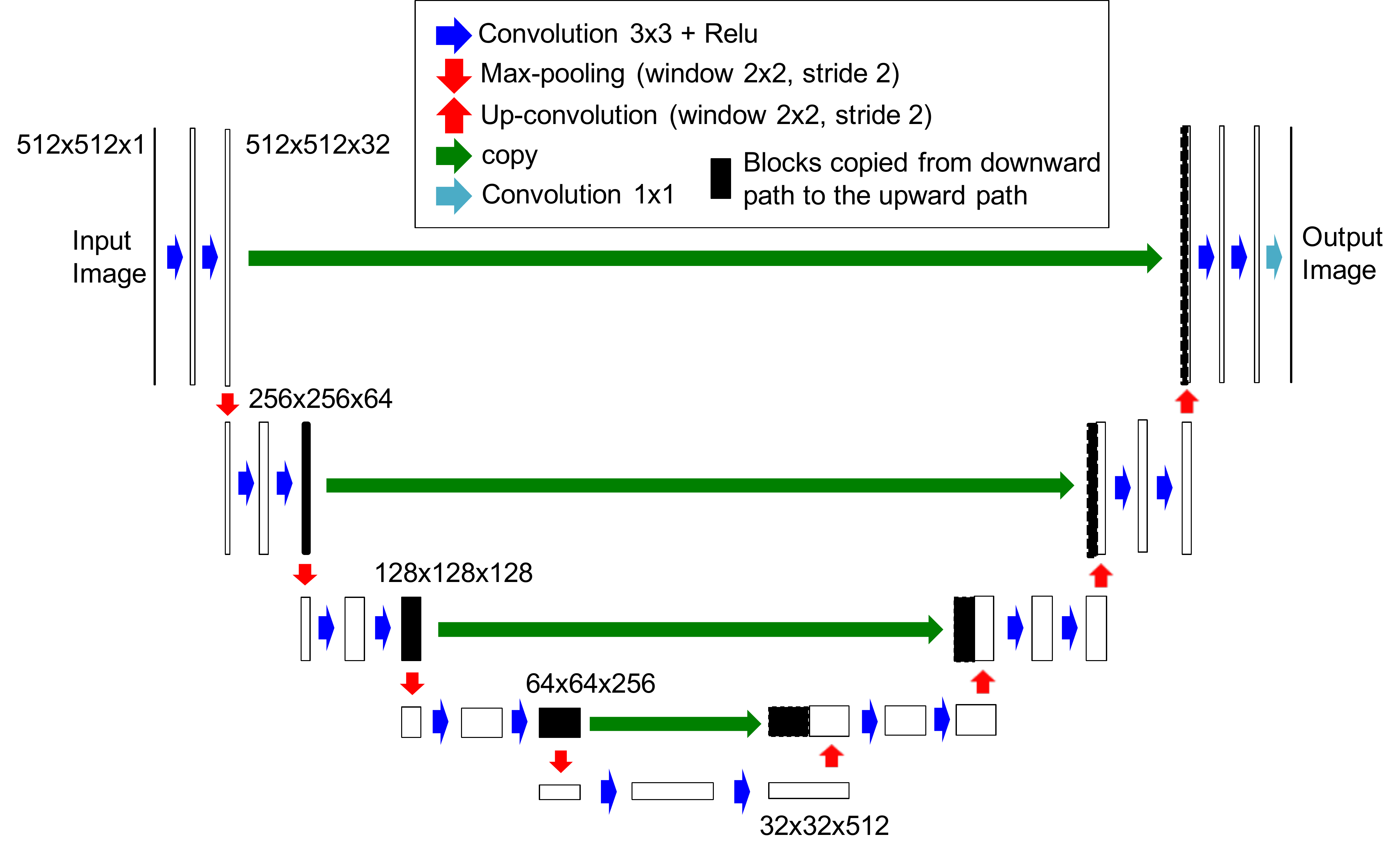}
\caption{Encoder-decoder network structure with skip connections (U-net), which wire the output of the first down-sampling layer to the input of the last up-sampling layer
and similarly for the second and the penultimate, etc. The two inputs to each up sampling layers are stacked as extra channels. The idea of having skip connections is to transfer the raw, non-abstract information, especially that of the high frequency signal, directly to the final output. Skip connections are also useful for mitigating the \emph{vanishing-gradient problem} and accelerating learning.}
\label{fig:unet}
\end{figure*}

Encoder-decoder networks refers to a particular structure of DNNs
which can be used for converting one input image into another.
An encoder-decoder network consists of multiple down-sampling layers with increasing numbers of channels (the encoder part),
followed by multiple up-sampling layers with decreasing numbers of channels (the decoder part).
The network has the form of an hourglass (Fig. \ref{fig:unet}).
Each down-sampling layer typically reduces the image size down to one quarter, by halving it in every dimension.
The number of channels is usually chosen to be twice that of the input to the layer,
so that the total information is compressed down to one half.
The purpose is to squeeze and abstract the information through the information bottleneck. The output of the middle layer is referred to as the \emph{latent feature}.
Each up-sampling layer then recovers the image size by quadrupling.

Let us consider how this architecture works.
Imagine that we have two persons at the ends of a phone line.
One person (the encoder) is given a plenty of 
noisy pictures $x_1,x_2,\ldots$, and the other (the decoder) is given the corresponding
clean pictures $y_1,y_2,\ldots$.
The encoder randomly chooses $x=x_i$ and tries to describe it over the phone.
The time is limited, and the person cannot say
``the left most corner has RGB value (30,41,200) and one pixel to the right has ...'',
but describes ``there is a large house in the middle with a blue-ish roof and ...''
The decoder, at the other end of the phone, tries to draw a picture $y$ based on the explanation by the encoder.
After finishing the drawing, 
the decoder is informed of the ID of the picture, which is $i$, so that 
the decoder can compare what is drawn, $y$, with the corresponding clean picture, $y_i$ (the \emph{ground truth}).
The comparison result is fed back to the two persons to improve;
the encoder improves the ability of explaining the contents of the noisy pictures, and the decoder improves drawing clean pictures from the encoder's explanation. 
In the end, they are supposed to master this game and become able to restore new noisy images 
without ground truth.
More specifically, their performance is evaluated by a \emph{reconstruction loss function}
$d(y,y_i)$ which measures the difference between two images, $y$ and $y_i$.
A typical choice for $d(y',y_i)$ is the $\ell^p$ distance between $y'$ and $y$
(recall that $y'$ and $y$ are just high-dimensional vectors).
In particular, the $\ell^2$ distance, which is the square root of the mean squared pixel-wise error, and the $\ell^1$ distance, which is the mean absolute pixel-wise error, 
are popularly used.

Through the above process, visual information on the pictures is abstracted to verbal information.
What is important is this abstract representation of the information.
A picture with a lot of noise, or with a missing part, can be recovered through this abstraction process.
Furthermore, if we train the decoder to draw in the style of Monet, 
a picture is first translated into words, and from the words the trained painter paints something like a Monet masterpiece.
This is the basic idea of image translation with use of encoder-decoder networks.

But how can we train the decoder to draw like Monet?
This is where the other key component, GANs, comes into the story.
If we have many pairs of photos corresponding to Monet's paintings, 
we can use the same way as above with this training dataset, by giving the photos to the encoder and the corresponding paintings to the decoder.
However, Monet never drew a smartphone, and the training dataset 
does not contain such a picture.
The chance is that the encoder and the decoder cannot do well 
when they are asked to handle a picture of a smartphone.
So we employ a third player called the \emph{discriminator} (or the \emph{critic}),
who will be trained to distinguish real Monet paintings from
forged ones by the decoder.
The encoder and the decoder now have two separate objectives;
they collaborate to convert a photo to a painting whose contents are the same as in the original photo,
and at the same time to deceive the third player. 
The team of the encoder and the decoder is called the \emph{generator}.
The discriminator is trained with real paintings of Monet and generated ones 
and has to tell how likely it is that a given picture was drawn by Monet.
They train themselves through friendly rivalry.
Their performance is evaluated in terms of the loss function.
Each player is associated with a tailored loss function which measures 
how well the player achieves his or her goal.
The loss function is designed so that the smaller its value, the better the player is doing.

The discriminator is usually a DNN with down sampling layers
whose objective and structure are the same as those used for classification tasks.
The generator is an encoder-decoder network 
whose objective is to convert an input image with which the discriminator 
makes a wrong classification, and at the same time the contents of 
 the original image are the same as those of the original image.
But how can we be assured that the contents of the original and the converted images are same?
We can think of two different situations.

First, a \emph{paired}-images dataset consists of
pairs of aligned images in the source domain $A$ and the target domain $B$.
For example, a photo of a cathedral and a painting by Monet of the cathedral with exactly the same angle make a paired image.
We want the generator $f$ to learn to convert 
$x\in A$ to $f(x)\in B$.
With a paired-images dataset, the discriminator will be trained to discriminate between a pair $(x,y)$ of a photo and the corresponding real painting
and a pair $(x,f(x))$ of a photo and the generated painting.
The discriminator's task is one of classification, and any classification loss can be used for training the discriminator.
The discriminator learns how Monet would paint based on a photo by optimizing the discriminator's loss.
This is called \emph{conditional GAN}, because the discriminator is shown the photo in addition to a forged or a real painting to judge its authenticity. 
The generator is given $x\in A$ and is trained to optimize
a weighted sum of the \emph{reconstruction loss}, measuring the closeness 
between $y$ and $f(x)$ and the \emph{adversarial loss}, which is the negative of the discriminator's loss for $(x,f(x))$.
This is an instance of \emph{supervised learning}.
The idea of using a conditional GAN for a general-purpose image-to-image translation has been extensively investigated in \emph{pix2pix} \cite{Isola2016}.
However, paired images are difficult to acquire in medical imaging, as the images must be aligned almost perfectly for making paired images.

\begin{figure*}\label{fig:cyclegan}
  \includegraphics[width=0.9\textwidth]{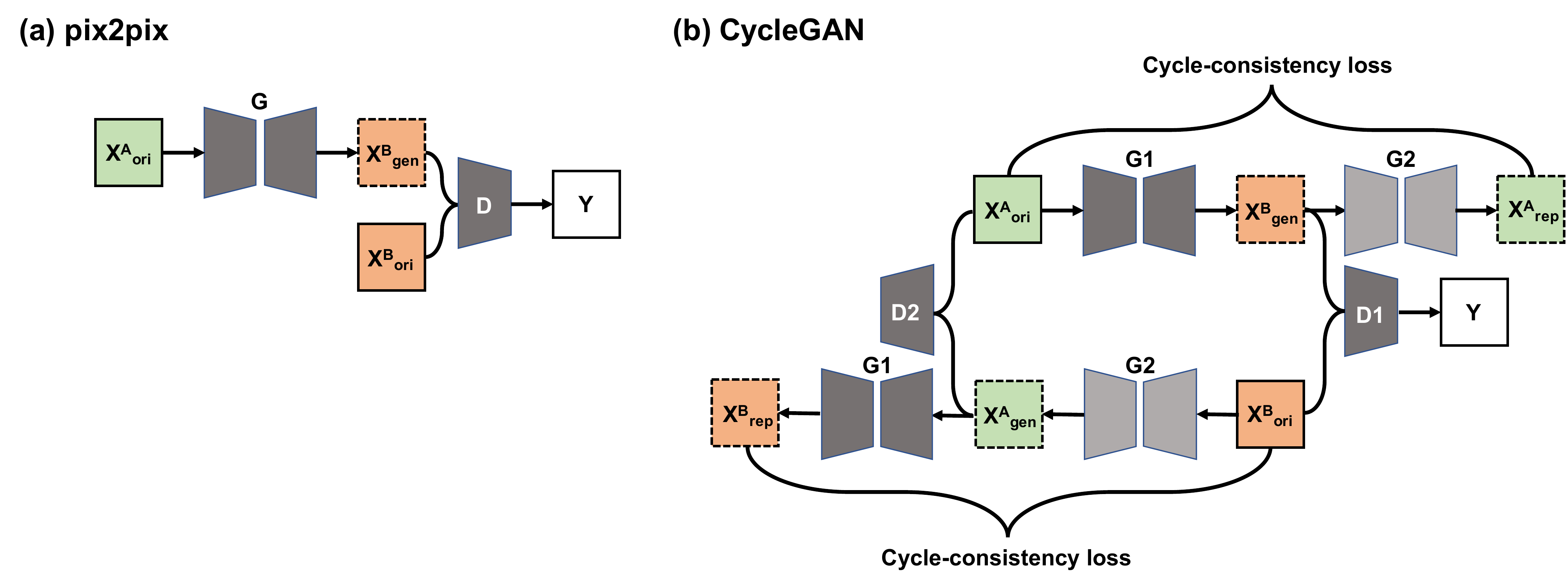}
\caption{The architecture of (a) pix2pix and (b) CycleGAN. (a) pix2pix requires perfectly aligned paired training images. A generator CNN is trained to generate images similar to images in domain $B$ from images in domain $A$, and discriminator CNN is trained simultaneously to distinguish the generated images from real images in domain $B$. Reconstruction loss measures how close by the real images in domain B and the generated images. On the other hand, (b) CycleGAN can learn a translation mapping in the absence of aligned paired images. The image generated from domain $A$ to domain $B$ by generator CNN (G1) is converted back to domain $A$ by another generator CNN (G2), and vice versa, in the attempt to optimize the cycle-consistency loss in addition to the adversarial loss.}
\label{fig:cyclegan}
\end{figure*}

On the other hand, an \emph{unpaired} image dataset consists of a set of images from the source domain $A$
and an independent set of images from the target domain $B$.
We do not know which image in $A$ corresponds to which in $B$.
Again, we want the generator $f$ to learn to convert 
$x\in A$ to $f(x)\in B$.
The discriminator is trained to distinguish real images $y$ from generated images $f(x)$. Note that, here, $y$ and $x$ are independent and do not correspond to each other; $y$ could be a painting of a waterlily, whereas $x$ is one of a smartphone.
In this case, it is difficult to preserve the contents under translation,
as the encoder and the decoder have no way to check if the decoder's drawing 
corresponds to what the encoder describes. 
That is, it is difficult to define a reconstruction loss as we do not 
have the ground truth on what $f(x)$ should be.
One trick to encourage the content preservation is to introduce another group of a generator $g$ and a discriminator, who will be trained to convert in the opposite direction from the target domain $B$ to the source domain $A$.
We demand that a image converted by the first group is converted back to the original, and vice versa.
That is, the \emph{cycle-consistency loss} $|x-g(f(x))|^2+|y-g(f(y))|^2$
should also be optimized.
This encourages both parties to learn one-to-one mappings.
The architecture of the whole system is depicted in Figure \ref{fig:cyclegan}.
This is an instance of \emph{unsupervised learning}.
The cycle-consistency loss is introduced in \emph{CycleGAN} \cite{CycleGAN2017}\footnote{Around the same time, very similar architectures such as UNIT,  
DiscoGAN, and DualGAN were introduced. Walender et al. \cite{Welander2018UNIT} evaluated UNIT  \cite{Liu2017UNIT} and CycleGAN for transformation between T1 and T2-weighted MRI images and showed that these two frameworks performed almost equally well.}.
This does not guarantee the preservation of the contents, however,
and we have to be careful when using CycleGAN.
Anatomic structures can be altered and non-existing tumors can be pretended
in the converted images.
Further to encourage structure preservation, 
the perceptual loss (or the content loss) \cite{artistic}
between the original and the converted image can be added as an optimization target of the generators.
The perceptual loss between two images $x$ and $y$ is defined to be
$|h(x)-h(y)|^2$, where $h$ is another DNN pretrained on natural images
(for example, the VGG19 network trained on the ImageNet dataset which is truncated at some layer is used for $h$).
The idea is to compare the high-level, abstract features of images extracted by a pretrained DNN;
DNNs trained with natural images are like human eyes.
The perceptual loss is generally useful when a direct pixel-by-pixel comparison does not make sense (e.g., when images contain strong noise).

Obviously, it is better to use pix2pix than CycleGAN when a paired image dataset is available
(see \cite{CycleGAN2017} for comparison study).
However, in case the alignment of paired images is not perfect, CycleGAN may perform better.

With this purely data-driven approach, any image-to-image translation tasks work almost in the same manner.
Only different sets of data consisting of task specific images are required.
In fact, AUTOMAP \cite{Zhu2018ImageRB} demonstrated the ubiquity of this approach in a medical setting by showing that the same network can be used for
reconstruction of PET, CT, and MRI.
AUTOMAP relies on an encoder-decoder network and does not use GANs.
What is special with reconstruction is that the spatial correspondence between the original and the reconstructed 
images is not straightforward.
The fundamental assumption behind convolutional neural networks is that
spatially close pixels are highly correlated, so that convolution with a small kernel such as $3\times 3$ captures meaningful features.
On the other hand, adjacent pixels, say, in the frequency domain as in MRI, do not necessarily affect 
neighboring pixels in the reconstructed image.
The novel idea of AUTOMAP is to use a few fully connected layers \emph{before} a typical convolutional encoder-decoder network.
The fully connected layers are supposed to learn mathematical transformations such as the (inverse) 
Fourier transformation and the Radon transformation, which require spatially non-local manipulation.

\begin{remark}
One popular variant of the encoder-decoder networks, called the \emph{U-Net} \cite{unet}, uses \emph{skip connections}
which wire the output of the first down-sampling layer to the input of the last up-sampling layer,
and similarly for the second and the penultimate, etc.
The two inputs to each up-sampling layer are stacked as extra channels.
The idea is to transfer the raw, non-abstract information, especially that of a high-frequency signal, directly to the final output.
Skip connections are also useful for mitigating the \emph{vanishing gradient problem} and accelerating learning.
\end{remark}

\begin{remark}
GANs are notoriously difficult to train, as a balance among different players is required.
Several techniques are known for a stable training,
such as the Wasserstain GAN with gradient penalty (WGAN-GP) \cite{WGAN-GP}, 
Progressive Growing GAN (PGGAN) \cite{PGGAN}, and spectral normalization \cite{spnorm}.
\end{remark}


\section{Noise reduction}
Noise reduction is the process of removing noise in images acquired through various imaging, typically those with low-dose CT (LDCT). 
This can be considered as translating of an image with noise to one with reduced noise.
Among model based approaches, iterative reconstruction (IR) with some kind of prior information has been used for the reduction of noise and artifact. However, its slow reconstruction speed and its poor output quality have limited its clinical application. Recently, denoising methods using deep neural networks (DNN) have been intensively studied (see \cite{denoiselist}). 
Most of the DNN-based methods involve post-processing of reconstructed images, which does not rely on raw projection data. 
For simplicity, we focus on the setting of denoising LDCT in order to obtain an image resembling normal-dose CT (NDCT).

Chen et al. \cite{Chen2017LC} proposed
a shallow encoder-decoder network trained with a paired dataset consisting of NDCT images and LDCT images. Here, NDCT images are acquired routinely, and the LDCT images are generated artificially by addition of Poisson noise to the corresponding NDCT images. The proposed DNN learns the end-to-end mapping from LDCT images to NDCT images. Once trained, the network takes LDCT images as input and converts it to one similar to NDCT images. The output quality is less plausible when it is compared with that of the state-of-the-art algorithms which we review later in this subsection. However, the paper of Chen et al. is simple and well-written and helps us to grasp the basic ideas on denoising with DNNs. 
Although this simple method greatly reduces the noise of LDCT images, a limitation is that the resulting images look over-smoothed, and sometimes lose structure details because these methods target minimizing only of reconstruction losses between NDCT images and converted NDCT-like images in the training dataset and they are not generalized well for new, unseen images.
One solution to this problem is the use of GAN-based methods such as pix2pix.
As we discussed in the previous section, pix2pix requires perfectly aligned paired training images, which it may be difficult to obtain in clinical settings.
There are two main ways of preparing such paired datasets.
\begin{itemize}
\item Noisy images are created by addition of artificial noise to high-quality NDCT images. The problem with this method is that the noise distribution is different from the real one.
\item Multiple acquisitions are performed at different CT radiation doses. 
The problem with this method is that the images are not perfectly aligned even if they are registered with DIR (deformable image registration).
\end{itemize}
With a paired dataset at hand, a generator DNN is trained to generate NDCT-like images from LDCT images
in the attempt to optimize a reconstruction loss
which measures how close the real NDCT images are the generated NDCT-like images.
A discriminator CNN is trained simultaneously to distinguish the generated NDCT-like images from real NDCT images. 
The generator tries to produce NDCT-like images which fool the discriminator
by optimizing the adversarial loss in addition to the reconstruction loss.
For a reconstruction loss, the squared sum of the pixel-wise difference in value (the squared $\ell^2$ loss) is often used,
but it results in blurry outputs.
Alternative loss functions including the perceptual feature loss \cite{perceptual}, \emph{structure loss} \cite{You2018}, and \emph{sharpness loss} \cite{YiBabyn2018} have been investigated.
Each of these loss functions is dependent on the accuracy of the alignment of paired images, and aligned paired training images are indispensable for using the pix2pix framework. However, in real clinical situations, it is difficult to obtain aligned paired LDCT and NDCT images. Creating LDCT by adding artificial noise to NDCT images involves the problem of a difference from the actual noise of LDCT, and multiple acquisitions at different doses cause problems of an additional radiation dose to patients and of positioning errors between acquisitions.

On the other hand, CycleGAN can learn a translation mapping in the absence of aligned paired images. Kang et al. \cite{Kang2019} applied CycleGAN to learn the mapping between the low- and normal-dose cardiac phases, which are not aligned exactly with each other due to the cardiac motion. 
Their method effectively reduces the noise in low-dose cardiac images while suppressing the deformation and loss of structures. 

\section{Super resolution}
The purpose of super-resolution (SR) is recovering a high-resolution image from a single low-resolution image. Most existing methods for SR learn mapping functions from external low- and high-resolution example pairs (paired datasets).
Conventional SR methods learn the dictionaries \cite{Timofte2013SR,Yang2008SR} or manifolds \cite{Bevilacqua2012SR,Chang2004SR} for modeling of the patch space. 
On the other hand, DNN-based methods learn an end-to-end mapping between low- and high-resolution images, thus implicitly achieving dictionaries or mapping functions for patch space by hidden layers of DNN. 
With SRCNN (Super-Resolution Convolutional Neural Network),
low-resolution input images are first upscaled to the desired size by use of bicubic interpolation and are then fed to an encoder-decoder network;
thus, end-to-end mapping between the bicubic upscaled version of a low-resolution image and a ground truth high-resolution image is learned. 
The SRCNN has been applied to mammography images \cite{Umehara2017SRmammo}, chest CT images \cite{Umehara2017SRChest}, and MRI images \cite{Plenge2012SRMRI}.

SRGAN (Super-Resolution Using a Generative Adversarial Network)  \cite{Ledig2016SRGAN} can be thought of as a GAN-fortified version of SRCNN.
An encoder-decoder network with more upsamling layers than downsampling layers is trained to recover detailed textures from heavily downsampled images,
and a discriminator is trained to differentiate between the super-resolved images and original high-quality images.
SRGAN has been applied to the generation of high-resolution brain MRI images from low-resolution images, and 3D convolution has been adopted for exploiting volumetric information \cite{Sanchez2018SRGANmri}.

\section{Image synthesis}

Medical image synthesis is performed for two purposes. 
One of these is to expand a small dataset with new, plausible examples for  training of DNNs for diagnosis and other tasks.
The other is to generate images virtually, images which are not acquired due to the clinical workflow or reduction of the acquisition time, cost, and dose. For the former purpose, transformations such as shifts, flips, zooms, or rotations have been performed traditionally. However, sufficient variations in the size, shape, and contrast of samples cannot be obtained, which results in the deterioration of the accuracy of the detection and classification task. To cope with this problem, unconditional synthesis by use of GAN, which generates images from noise without any other conditional information, has been performed for generating various plausible training images \cite{Chuquicusma2017DCGAN,Adar2018DCGAN,Bermudez2018DCGAN,Madani2018DCGAN,Korkinof2018PGGAN}. For the latter purpose, cross modality synthesis such as generating CT-like images from MRI images by use of GAN has been proposed \cite{Wolterink2017CycleGAN,Hiasa2018CycleGAN,Zhang2018CycleGAN}. The CT-like images synthesized from MRI images may be applied to dose calculation (treatment planning) for upcoming MRI-only radiotherapy. 

\subsection{Unconditional synthesis} 
Unconditional synthesis by use of GAN generates images from noise without any other conditional information such as the boundary of structures. This can be considered as image-to-image translation in which the input images are just randomly generated noise.
Deep convolutional GAN (DCGAN) and its improved variant, Progressive Growing of GAN (PGGAN), have been used for generation of medical images. DCGAN has been used for generation of synthetic samples of CT images of lung nodules \cite{Chuquicusma2017DCGAN} and liver lesions \cite{Adar2018DCGAN}, MRI images of the brain \cite{Bermudez2018DCGAN}, and X-ray images of the chest \cite{Madani2018DCGAN}. PGGAN has been used for generating synthetic samples of mammography images \cite{Korkinof2018PGGAN}. DCGAN has been applied to image resolution of up to $256\times 256$, but PGGAN can be applied to an image resolution of up to $1024\times 1024$. The key idea of PGGAN is to grow both the generator and the discriminator progressively: starting from a low resolution, layers that model increasingly fine details as training progresses are added. 

These GAN-based augmented images were found to be beneficial for various classification and segmentation tasks when combined with real images. Wu et al. \cite{Wu2018GANaug} reported that a GAN-based augmentation improved the area under the curve (AUC) of a mammogram patch-based classification by 0.009 $(0.887 \to 0.896)$  over a traditional augmentation approach. Mok et al. \cite{Mok2019GANaug} reported that a PGGAN-based augmentation improved the dice coefficient of segmentation of a brain tumor by 0.03 $(0.81 \to 0.84)$ over a traditional augmentation approach. Frid-Adar et al. \cite{FridAdar2018GANaug} reported that a DCGAN-based augmentation improved the classification of a liver lesion by $7.1\%$ $(78.6 \% \to 85.7\%)$ in sensitivity and $4.0\%$ $(88.4\% \to 92.4\%)$ in specifity over a traditional augmentation approach. 

\subsection{Cross modality synthesis}
Cross-modality synthesis by use of encoder-decoder networks has been explored in several studies, which included MRI-to-CT synthesis \cite{Nie2016EDN,Han2017EDN} and cone beam CT-to-planning CT synthesis \cite{Kida2017EDN}.
In cases where images from two different modalities can be spatially aligned correctly, such as PET-CT, where PET images and CT images are acquired simultaneously \cite{Avi2019pix2pix}, pix2pix can be used.
On the other hand, CycleGAN is used in case it is clinically difficult or impossible to acquire aligned paired training images from two different modalities such as CT and MRI images, which are acquired with different modality \cite{Wolterink2017CycleGAN,Hiasa2018CycleGAN,Zhang2018CycleGAN}.

Wolterink et al. \cite{Wolterink2017CycleGAN} synthesized CT images of the head from MRI images of the head by using CycleGAN, and they showed that CycleGAN using unpaired images created CT images that looked more realistic and contained fewer artifacts and blurring than did GAN combined with voxel-wise loss by use of aligned images. The difference could be due to misalignment between MRI and CT images, which is ignored in training with unpaired images by use of CycleGAN. However, CycleGAN sometimes fails to preserve the boundaries of structures in images such as those of the pelvic region, which has large variations in the anatomic structures between images from two different modalities due to the presence of joints, muscles, intestines, and rectum. Hiasa et al. \cite{Hiasa2018CycleGAN} extended CycleGAN by incorporating a loss function named gradient consistency loss, which evaluates consistency of image gradient at each pixel between the original and the synthesized images to preserve the boundary of structures in the pelvic region at MRI-to-CT synthesis. They showed that synthesized images with gradient consistency loss preserved the shape near the femoral head and adductor muscles better than did those without gradient consistency loss. Kida et al. \cite{Kida2019CycleGAN} also extended CycleGAN by incorporating several losses for better preservation of the boundary of structures in the pelvic region in cone beam CT-to-planning CT synthesis. They showed that edge sharpness and structures were preserved not only in large, bulky tissues such as the rectum and bladder, but also in small isolated structures such as the small intestine and intestinal gas. They showed, with some initial weightings of the neural network, that the generators completely failed to learn a good mapping and produced totally distorted images. This kind of failure tends not to be reported, but practitioners have to be warned when they plan to deploy DNN-based methods.
Zhang et al. \cite{Zhang2018CycleGAN} employed shape consistency loss that is produced by another image-to-image translation DNN for segmentation, named the segmentor.
One segmentor is introduced in each domain so that it can learn segmentation, in addition to the generator and the discriminator.
The segmentor's outputs for the original and the translated images are compared in the shape consistency loss, which is a type of perceptual loss.
During training, the segmentor takes advantage of the generator by using synthetic images as augmented data,
and the generator and the segmentor prompt each other to preserve the anatomic structures under translation through the loss functions. 
Finally, the segmentors provide implicit shape constraints on the anatomic structures during image synthesis.
It should be noted that semantic labels (annotated segmentation) of training images from both modalities are required. 

\section{Reconstruction}\label{sec:reconst}
Imaging such as CT, MRI, and PET can also be thought of as a specific instance of image-to-image translation
from a volumetric object (density distribution) in three-dimensional space
to a (series of) projection image(s).
In other words, imaging is a mapping
\[
Im: \R^{d_1\times h_1 \times w_1} \to 
\R^{d_2\times h_2 \times w_2}. 
\]
Reconstruction is then the translation from the projection image back to
the original distribution and is regarded as a mapping
\[
Rec: \R^{d_2\times h_2 \times w_2} \to 
\R^{d_1\times h_1 \times w_1},
\]
so that the composition $Rec\circ Im$ is the identity map.
In the ideal setting of no noise and infinite resolution, 
mathematics allows us to find a good reconstruction mapping $Rec$ for each imaging mapping $Im$,
such as the filtered back-projection for fan-beam CT.
However, the reality is that we have to tackle technical difficulties such as 
noise and low resolution due to scanner and dose limitations, as well as numerical error of the algorithm.
In this section, we discuss what machine learning can offer for improved reconstruction.

\subsection{Hybrid approach}\label{sec:hybrid}
As reconstruction can be thought of as image-to-image translation,
the DNNs discussed in \S \ref{sec:AEN} (e.g., AUTOMAP \cite{Zhu2018ImageRB}) can be directly applicable
if we have enough data consisting of pairs of original and translated images.
That is, we can approximate the reconstruction mapping by a DNN.
Purely data-driven methods do not rely on the fact that imaging is the result of a physical process that is subject to certain rules. 
Here, we see how we can exploit knowledge from physics and mathematics and combine it
with data-driven approaches to enhance the performance of reconstruction.
In this way, we will achieve a better quality of reconstruction with a lesser amount of training image pairs.

Most imaging techniques guarantee at least theoretically that for each projection there exists a unique distribution so that the projection and the original distribution contain exactly the same amount of information. 
However, due to the noise and the finite resolution, some information gets lost in the imaging process,
which makes the reconstruction \emph{ill-posed} and difficult.
That is, given a projection image, we cannot expect to find one and only one distribution which projects to it.
Therefore, the aim of reconstruction is to find a distribution which translates as closely as possible to the given projection
with respect to some goodness and closeness criteria.

In practice, we (assume to) know the mathematical model of the imaging, which is represented by 
\begin{equation}\label{eq:img}
y = Im(x),
\end{equation}
where $x$ is a distribution, and $y$ is the corresponding projection. 
The imaging mapping $Im$ is often called the \emph{forward operator} and is determined by
the configuration of the imaging equipment such as the scanner geometry and the sub-sampling scheme.
If $Im$ were invertible, we could directly solve $x=Im^{-1}(y)$, 
but usually equation \eqref{eq:img} is ill-posed 
due to hardware limitations, and there exists no $Im^{-1}$.
Therefore, reconstruction takes the form of an optimization problem;
namely, we would like to find an $x^*$ which minimizes the following
\begin{equation}\label{eq:cost}
L(x) = d(y,Im(x)) + \lambda E(x),
\end{equation}
where the \emph{reconstruction loss term} (or the \emph{data fidelity term}) $d(y,Im(x))$ shows how close $y$ and $Im(x)$ are, 
the \emph{regularization term} $E(x)$ shows how appropriate
 $x$ is as a reconstructed image, and $\lambda> 0$ determines the balance between the two criteria.
 Such an $x^*$ is often denoted by $\mathrm{argmin}(L(x))$.
There are many choices for $d$, $E$, and $\lambda$. 
Once we fix $Im$, $d$, $E$, and $\lambda$, all we have to do for the reconstruction 
is simply optimizing (finding an $x$ which attains the small value of) \eqref{eq:cost}.
The reader may notice that this is exactly the same as the formulation of iterative reconstruction.

Popular choices for $d$ are the $\ell^1$ norm and the $\ell^2$ norm.
Both are standard distances in Euclidean space.
Whereas $\ell^2$ greatly penalizes large deviation in values, 
it permits a small deviation in a large region. 
These characteristics often lead to a blurry output.
On the other hand, $\ell^1$ prefers two images to be exactly the same at most of pixels,
but allows the rest of pixels to deviate a lot.
It is wise to keep these characteristics in mind to design loss functions tailored for specific applications.
We can also use a DNN for defining $d$ which measures an abstract distance between images such as a perceptual loss (see \S \ref{sec:AEN}).

The purpose of the regularization term is threefold:
\begin{enumerate}
    \item to incorporate prior about images for a higher quality output
    \item to suppress over-fitting for a better generalization
    \item to mitigate ill-posedness for better convergence of the optimization 
\end{enumerate}
For example, 
it is reasonable to ask for a solution image which has some spatial uniformity.
A popular conventional choice is the $\ell^1$ norm of the image gradient (the total variation).
We can also use DNNs to learn image priors from data \cite{Chang_2017_ICCV}.
This idea is similar to GANs; a DNN is trained with the images in the reconstructed domain
to learn their distribution.
In fact, we can use the adversarial loss from a discriminator trained with high-quality reconstructed images
and images $x$ which are in the middle of reconstruction.
One problem, however, is that such high-quality images may not be available, as we have no way of knowing the ground-truth distribution of our body except for phantoms.
Therefore, using a DNN for image priors is useful when high-quality reconstruction images (e.g., planning CT)
are available and we would like to do the reconstruction with limited-quality projections (low-dose CT). 
The regularization term $E$ can consist of many sub-terms. For example, 
we can take $E$ to be a weighted sum of the total variation and the adversarial loss.


Another interesting way of regularisation, called the deep image prior, is 
proposed in \cite{deepimageprior}.
This does not take the form of the added term $E$.
Instead, the idea is to replace $x$ with $f_w(0)$ in \eqref{eq:cost},
where $f$ is a DNN with weight $w$.
The optimization is performed not directly on $x$, but indirectly on $w$.
One way to understand this is to think of 
DNNs as restricted approximators of arbitrary signals, just as
trigonometric functions and wavelet functions are conventionally used for approximating signals.
We know from sparse modeling that, if we restrict ourselves to use only 
a small number of function bases, we obtain better results.
As DNNs are suitable for representing visual information which our eyes can perceive,
it may be more reasonable to use DNNs rather than trigonometric functions and wavelets as approximators.
In other words, whereas $x$ can take any value in the entire Euclidean space, 
$f_w(0)$ is constrained within the image of a mapping $w \mapsto f_w(0)$,
which is more likely to be visually plausible.

In any case, the motto of regularization is 
``do not give complete freedom, but restrict in a reasonable manner'',
as in ordinary education.

Both model-based and data-driven approaches can be integrated
in the form of the optimization of \eqref{eq:cost}.
This hybrid approach has been studied actively in recent years 
(see, for example, \cite{Chang_2017_ICCV,Adler_2017,Lucas2018}).

Image-to-image translation by DNNs as reviewed in the previous sections can be seen as replacing
the whole process of the optimization of \eqref{eq:cost} by DNNs
which directly find $x_*$ from given inputs $y$.

\section{Sample codes}\label{sec:sample}
In this section, we give a hands-on account on what we encounter in the deployment of DNNs for a specific task.
We use our codes for image-to-image translation for both a paired and an unpaired dataset,
which would cover most of the topics in this article except for those in \S\ref{sec:hybrid}.
Our codes are written in Python, with use of a deep learning framework, Chainer \cite{chainer}.
For demonstration purposes, 
we work on a simple denoising scenario, where we assume that a high-quality training dataset is available.
For other image-to-image translation tasks, the workflow is more or less the same except for the way in which datasets are acquired.
Note that our purpose in the section is to give an overview of how an actual procedure may look, 
and we do not mean to make a practical denoising model.

For DNNs, we use the code available at \cite{github2} for 
general image-to-image translation DNNs for paired datasets.
How to set up the running environment 
and what to type in to run the code are given in the code's documentation.

For datasets, we use CPTAC-SAR \cite{CPTAC-SAR}, which is made publicly available at The Cancer Imaging Archive. 
The dataset contains normal-dose CT images of the Sarcomas cohort in the DICOM format.
We make low-quality noisy CT images by adding artificial Poisson noise.
This is done by the Python script included in \cite{github2}, as instructed in the documentation.
Now, we have hundreds of pairs of a noisy and a clean CT image,
and we train a DNN based on pix2pix to learn a mapping from a noisy image to a clean one.
During the training, the dataset is \emph{augmented} on the fly by random crop and flip, and by addition of random Gaussian noise.
There are numerous \emph{hyper-parameters} for training;
\begin{itemize}
\item how long (how many \emph{epochs}) the training lasts; recall that training is an optimization, and the user has to decide when to stop.
\item how many layers the encoder-decoder network and the discriminator network have; network architectures have to be fixed
\item how to weight different loss functions,
\end{itemize}
just to mention a few. 
The code comes with reasonable default values for the hyper-parameters, but to get the best performance, the user has to \emph{tune}
them depending on the dataset and the goal of the task. 
Basically, we search by trial and error to find the best hyper-parameter setting.
One training takes from hours to days, so that tuning can be very time-consuming.

Once we obtain a well-trained model, we can use it to denoise new DICOM images.
The model can process dozens of images per second even without a GPU.
The trained model may or may not \emph{generalize} to other CT images.
DNNs basically learn how to handle images which are similar to those in the training dataset.
New images can have totally different noise characteristics, or contain different organs.
When the trained model's performance is not satisfactory, we have to find other datasets and/or 
modify the network architecture, and/or tune hyper-parameters until we are satisfied.

This is a typical procedure of deploying DNNs in an image processing task.
We can also use another code \cite{github1} to work with an unpaired dataset,
e.g., with noisy images from one patient and clean images from another.

\section{Conclusion}
In recent years, the use of DNNs has been becoming dominant in medical image processing.
Many tasks are considered as instances of image-to-image translation.
Two key inventions, encoder-decoder netoworks and GANs, have improved the usability of DNNs in image-to-image translation,
and there are two fundamental architectures that utilize them; pix2pix and CycleGAN.
They are used for different types of datasets;
pix2pix is preferable when a dataset consisting of paired images is available,
whereas CycleGAN can work with a dataset consisting of unpaired images.
The following table summarizes various medical applications that we reviewed in this paper in terms of their base architectures
(Table \ref{tab:table}).

\begin{table*}[ht]\center
\scalebox{0.82}{
\begin{tabular}{|>{\columncolor[gray]{0.8}}c|cccccccc|}
  \hline
\rowcolor{LightCyan}
  \hspace{1cm} & Encoreder-decorder & pix2pix & CycleGAN & UNIT & SRCNN & SRGAN & DCGAN & PGGAN \\
  \hline
  Noise reduction & \cite{Chen2017LC} & \cite{perceptual} \cite{You2018} \cite{YiBabyn2018} & \cite{Kang2019} & \hspace{1cm} & \hspace{1cm} & \hspace{1cm} & \hspace{1cm} & \hspace{1cm}\\ 
  \hline
  Super-resolution & \hspace{1cm} & \hspace{1cm} & \hspace{1cm} & \hspace{1cm} & \cite{Umehara2017SRmammo} \cite{Umehara2017SRChest} \cite{Plenge2012SRMRI} & \cite{Ledig2016SRGAN} \cite{Sanchez2018SRGANmri} & \hspace{1cm} & \hspace{1cm} \\
  \hline
  Unconditional synthesis & \hspace{1cm} & \hspace{1cm} & \hspace{1cm} & \hspace{1cm} & \hspace{1cm} & \hspace{1cm} & \cite{Chuquicusma2017DCGAN} \cite{Adar2018DCGAN} \cite{Bermudez2018DCGAN} \cite{Madani2018DCGAN} \cite{FridAdar2018GANaug} & \cite{Korkinof2018PGGAN} \cite{Mok2019GANaug} \\
  \hline
  Cross modality synthesis & \cite{Nie2016EDN} \cite{Han2017EDN} \cite{Kida2017EDN} & \cite{Avi2019pix2pix} & \cite{Wolterink2017CycleGAN} \cite{Hiasa2018CycleGAN} \cite{Kida2019CycleGAN} \cite{Zhang2018CycleGAN} \cite{Welander2018UNIT} & \cite{Liu2017UNIT} \cite{Welander2018UNIT} & \hspace{1cm} & \hspace{1cm} & \hspace{1cm} & \hspace{1cm}\\
  \hline
  \end{tabular}
  }
  \caption{A brief summary of tasks and networks in the reviewed publications.}\label{tab:table}
\end{table*}

In some cases, conventional model-based approaches do better than DNNs, especially when the available data are limited.
Also, data-driven approaches often lack explainability, and we may get totally wrong outputs for certain inputs.
Thus, quantifying uncertainty and establishing evaluation and validation schemes
is crucial for the reliable use in clinical applications.
Explainability and quantifying uncertainty are among the high-priority research topics in machine learning (e.g., \cite{Tanno17}).

We would like to emphasize that we have to be careful when using DNN-based methods.
However, at the same time, we must not be too afraid of using DNNs.
After all, they are merely tools.
As long as the user knows their proper usage and limitations, they are an extreme powerful companion for practitioners in clinical medicine.
They are like spelling helpers for authors.
We have written this article with a word processor that has a spelling helper.
It corrects spelling automatically, but sometimes it makes mistakes.
Nevertheless, the number of corrected words is much larger than that of mis-corrected words.
We just should not trust it completely, but have to check the words again by ourselves.

For each medical image-processing scenario, 
the best method should be a hybrid combining the advantages of both data-driven and model-based approaches,
utilizing domain-specific knowledge to DNNs.
Some programing frameworks have been developed that enable hybrid approaches relatively easily.
For example, ODL \cite{ODL} is a Python-based framework for reconstruction algorithms 
which is integrated with deep-learning frameworks.
For innovative advancement,
radiologists, physicists, mathematicians, and computer scientists
have to team up and conduct interdisciplinary research.

\section*{Compliance with Ethical Standards}
Funding: Kaji was partially supported by JST PRESTO, Grant Number JPMJPR16E3 Japan.

Conflict of Interest:  The authors declare that they have no conflict of interest.

Ethical approval: This article does not contain any studies with human participants or animals performed by any of the authors.

\section*{Acknowledgement}
We thank Mrs. Lanzl at Chicago University for her English proofreading.

\bibliographystyle{ieeetr}
\bibliography{ref.bib}

\begin{thebibliography}{10}

\bibitem{Sahiner2018DeepLI}
B.~Sahiner, A.~Pezeshk, L.~M. Hadjiiski, X.~Wang, K.~Drukker, K.~H. Cha, R.~M.
  Summers, and M.~L. Giger, ``Deep learning in medical imaging and radiation
  therapy.,'' {\em Medical physics}, 2018.

\bibitem{github1}
S.~Kaji, ``Image translation by {CNNs} trained on unpaired data.''
  \url{https://github.com/shizuo-kaji/UnpairedImageTranslation}, 2019.

\bibitem{github2}
S.~Kaji, ``Image translation for paired image datasets (automap + pix2pix).''
  \url{https://github.com/shizuo-kaji/PairedImageTranslation}, 2019.

\bibitem{DBLP:series/acvpr/978-3-319-42998-4}
L.~Lu, Y.~Zheng, G.~Carneiro, and L.~Yang, eds., {\em Deep Learning and
  Convolutional Neural Networks for Medical Image Computing - Precision
  Medicine, High Performance and Large-Scale Datasets}.
\newblock Advances in Computer Vision and Pattern Recognition, Springer, 2017.

\bibitem{DBLP:conf/miccai/2018mlmir}
F.~Knoll, A.~K. Maier, and D.~Rueckert, eds., {\em Machine Learning for Medical
  Image Reconstruction - First International Workshop, {MLMIR} 2018, Held in
  Conjunction with {MICCAI} 2018, Granada, Spain, September 16, 2018,
  Proceedings}, vol.~11074 of {\em Lecture Notes in Computer Science},
  Springer, 2018.

\bibitem{Litjens2017ASO}
G.~J.~S. Litjens, T.~Kooi, B.~E. Bejnordi, A.~A.~A. Setio, F.~Ciompi,
  M.~Ghafoorian, J.~van~der Laak, B.~van Ginneken, and C.~I. S{\'a}nchez, ``A
  survey on deep learning in medical image analysis,'' {\em Medical image
  analysis}, vol.~42, pp.~60--88, 2017.

\bibitem{fiji}
J.~Schindelin, I.~Arganda-Carreras, E.~Frise, V.~Kaynig, M.~Longair,
  T.~Pietzsch, S.~Preibisch, C.~Rueden, S.~Saalfeld, B.~Schmid, J.-Y. Tinevez,
  D.~J. White, V.~Hartenstein, K.~Eliceiri, P.~Tomancak, and A.~Cardona,
  ``Fiji: an open-source platform for biological-image analysis,'' {\em Nat
  Meth}, vol.~9, pp.~676--682, July 2012.

\bibitem{nielsenneural}
M.~A. Nielsen, ``Neural networks and deep learning,'' 2018.

\bibitem{GoodBengCour16}
I.~J. Goodfellow, Y.~Bengio, and A.~Courville, {\em Deep Learning}.
\newblock Cambridge, MA, USA: MIT Press, 2016.
\newblock \url{http://www.deeplearningbook.org}.

\bibitem{Safran2017}
I.~Safran and O.~Shamir, ``Depth-width tradeoffs in approximating natural
  functions with neural networks,'' in {\em Proceedings of the 34th
  International Conference on Machine Learning - Volume 70}, ICML'17,
  pp.~2979--2987, JMLR.org, 2017.

\bibitem{universal}
F.~Scarselli and A.~C. Tsoi, ``Universal approximation using feedforward neural
  networks: A survey of some existing methods, and some new results,'' {\em
  Neural Netw.}, vol.~11, pp.~15--37, Jan. 1998.

\bibitem{NIPS2017_7203}
Z.~Lu, H.~Pu, F.~Wang, Z.~Hu, and L.~Wang, ``The expressive power of neural
  networks: A view from the width,'' in {\em Advances in Neural Information
  Processing Systems 30} (I.~Guyon, U.~V. Luxburg, S.~Bengio, H.~Wallach,
  R.~Fergus, S.~Vishwanathan, and R.~Garnett, eds.), pp.~6231--6239, Curran
  Associates, Inc., 2017.

\bibitem{dumoulin2016guide}
V.~{Dumoulin} and F.~{Visin}, ``{A guide to convolution arithmetic for deep
  learning},'' 2016.
\newblock arXiv:1603.07285.

\bibitem{pixelshuffler}
W.~Shi, J.~Caballero, F.~Huszar, J.~Totz, A.~P. Aitken, R.~Bishop, D.~Rueckert,
  and Z.~Wang, ``Real-time single image and video super-resolution using an
  efficient sub-pixel convolutional neural network,'' in {\em 2016 {IEEE}
  Conference on Computer Vision and Pattern Recognition, {CVPR} 2016, Las
  Vegas, NV, USA, June 27-30, 2016}, pp.~1874--1883, 2016.

\bibitem{deconvnoise}
A.~Odena, V.~Dumoulin, and C.~Olah, ``Deconvolution and checkerboard
  artifacts,'' {\em Distill}, 2016.

\bibitem{batchnorm}
S.~Ioffe and C.~Szegedy, ``Batch normalization: Accelerating deep network
  training by reducing internal covariate shift.,'' in {\em ICML} (F.~R. Bach
  and D.~M. Blei, eds.), vol.~37 of {\em JMLR Workshop and Conference
  Proceedings}, pp.~448--456, JMLR.org, 2015.

\bibitem{instancenorm}
D.~Ulyanov, A.~Vedaldi, and V.~S. Lempitsky, ``Improved texture networks:
  Maximizing quality and diversity in feed-forward stylization and texture
  synthesis,'' in {\em 2017 {IEEE} Conference on Computer Vision and Pattern
  Recognition, {CVPR} 2017, Honolulu, HI, USA, July 21-26, 2017},
  pp.~4105--4113, 2017.

\bibitem{Fredrikson2015}
M.~Fredrikson, S.~Jha, and T.~Ristenpart, ``Model inversion attacks that
  exploit confidence information and basic countermeasures,'' in {\em
  Proceedings of the 22Nd ACM SIGSAC Conference on Computer and Communications
  Security}, CCS '15, (New York, NY, USA), pp.~1322--1333, ACM, 2015.

\bibitem{unet}
O.~Ronneberger, P.~Fischer, and T.~Brox, ``U-net: Convolutional networks for
  biomedical image segmentation,'' in {\em Medical Image Computing and
  Computer-Assisted Intervention - {MICCAI} 2015 - 18th International
  Conference Munich, Germany, October 5 - 9, 2015, Proceedings, Part {III}},
  pp.~234--241, 2015.

\bibitem{GAN}
I.~Goodfellow, J.~Pouget-Abadie, M.~Mirza, B.~Xu, D.~Warde-Farley, S.~Ozair,
  A.~Courville, and Y.~Bengio, ``Generative adversarial nets,'' in {\em
  Advances in Neural Information Processing Systems 27} (Z.~Ghahramani,
  M.~Welling, C.~Cortes, N.~D. Lawrence, and K.~Q. Weinberger, eds.),
  pp.~2672--2680, Curran Associates, Inc., 2014.

\bibitem{ganlist}
X.~Yi, ``Awesome {GAN} for medical imaging.''
  \url{https://github.com/xinario/awesome-gan-for-medical-imaging}, 2019.

\bibitem{Isola2016}
P.~Isola, J.~Zhu, T.~Zhou, and A.~A. Efros, ``Image-to-image translation with
  conditional adversarial networks,'' in {\em 2017 {IEEE} Conference on
  Computer Vision and Pattern Recognition, {CVPR} 2017, Honolulu, HI, USA, July
  21-26, 2017}, pp.~5967--5976, 2017.

\bibitem{CycleGAN2017}
J.~Zhu, T.~Park, P.~Isola, and A.~A. Efros, ``Unpaired image-to-image
  translation using cycle-consistent adversarial networks,'' in {\em {IEEE}
  International Conference on Computer Vision, {ICCV} 2017, Venice, Italy,
  October 22-29, 2017}, pp.~2242--2251, 2017.

\bibitem{Welander2018UNIT}
P.~Welander, S.~Karlsson, and A.~Eklund, ``Generative adversarial networks for
  image-to-image translation on multi-contrast {MR} images - {A} comparison of
  {CycleGAN} and {UNIT},'' 2018.
\newblock arXiv:1806.07777.

\bibitem{Liu2017UNIT}
M.-Y. Liu, T.~Breuel, and J.~Kautz, ``Unsupervised image-to-image translation
  networks,'' in {\em Advances in Neural Information Processing Systems 30}
  (I.~Guyon, U.~V. Luxburg, S.~Bengio, H.~Wallach, R.~Fergus, S.~Vishwanathan,
  and R.~Garnett, eds.), pp.~700--708, Curran Associates, Inc., 2017.

\bibitem{artistic}
L.~A. {Gatys}, A.~S. {Ecker}, and M.~{Bethge}, ``Image style transfer using
  convolutional neural networks,'' in {\em 2016 IEEE Conference on Computer
  Vision and Pattern Recognition (CVPR)}, pp.~2414--2423, June 2016.

\bibitem{Zhu2018ImageRB}
B.~O. Zhu, J.~Z. Liu, B.~R. Rosen, and M.~S. Rosen, ``Image reconstruction by
  domain-transform manifold learning,'' {\em Nature}, vol.~555, pp.~487--492,
  2018.

\bibitem{WGAN-GP}
I.~Gulrajani, F.~Ahmed, M.~Arjovsky, V.~Dumoulin, and A.~C. Courville,
  ``Improved training of {W}asserstein {GANs},'' in {\em Advances in Neural
  Information Processing Systems 30} (I.~Guyon, U.~V. Luxburg, S.~Bengio,
  H.~Wallach, R.~Fergus, S.~Vishwanathan, and R.~Garnett, eds.),
  pp.~5767--5777, Curran Associates, Inc., 2017.

\bibitem{PGGAN}
T.~Karras, T.~Aila, S.~Laine, and J.~Lehtinen, ``Progressive growing of {GANs}
  for improved quality, stability, and variation,'' in {\em 6th International
  Conference on Learning Representations, {ICLR} 2018, Vancouver, BC, Canada,
  April 30 - May 3, 2018, Conference Track Proceedings}, 2018.

\bibitem{spnorm}
T.~Miyato, T.~Kataoka, M.~Koyama, and Y.~Yoshida, ``Spectral normalization for
  generative adversarial networks,'' in {\em 6th International Conference on
  Learning Representations, {ICLR} 2018, Vancouver, BC, Canada, April 30 - May
  3, 2018, Conference Track Proceedings}, 2018.

\bibitem{denoiselist}
J.~Sinyu, ``{CT} image denoising with deep learning.''
  \url{https://github.com/SSinyu/CT_DENOISING_REVIEW}, 2018.

\bibitem{Chen2017LC}
H.~Chen, Y.~Zhang, W.~Zhang, P.~Liao, K.~Li, J.~Zhou, and G.~Wang, ``Low-dose
  {CT} via convolutional neural network,'' {\em Biomed. Opt. Express}, vol.~8,
  pp.~679--694, Feb 2017.

\bibitem{perceptual}
Q.~Yang, P.~Yan, Y.~Zhang, H.~Yu, Y.~Shi, X.~Mou, M.~K. Kalra, Y.~Zhang,
  L.~Sun, and G.~Wang, ``Low-dose {CT} image denoising using a generative
  adversarial network with {W}asserstein distance and perceptual loss,'' {\em
  IEEE Transactions on Medical Imaging}, vol.~37, pp.~1348--1357, June 2018.

\bibitem{You2018}
C.~You, Q.~Yang, H.~Shan, L.~Gjesteby, G.~Li, S.~Ju, Z.~Zhang, Z.~Zhao,
  Y.~Zhang, W.~Cong, and G.~Wang, ``Structurally-sensitive multi-scale deep
  neural network for low-dose {CT} denoising,'' {\em {IEEE} Access}, vol.~6,
  pp.~41839--41855, 2018.

\bibitem{YiBabyn2018}
X.~Yi and P.~Babyn, ``Sharpness-aware low-dose {CT} denoising using conditional
  generative adversarial network,'' {\em Journal of Digital Imaging}, vol.~31,
  pp.~655--669, Oct 2018.

\bibitem{Kang2019}
E.~{Kang}, H.~J. {Koo}, D.~H. {Yang}, J.~B. {Seo}, and J.~C. {Ye},
  ``{Cycle-consistent adversarial denoising network for multiphase coronary CT
  angiography},'' {\em Medical Physics}, vol.~46, pp.~550--562, Feb. 2019.

\bibitem{Timofte2013SR}
R.~Timofte, V.~D. Smet, and L.~V. Gool, ``Anchored neighborhood regression for
  fast example-based super-resolution,'' {\em 2013 IEEE International
  Conference on Computer Vision}, pp.~1920--1927, 2013.

\bibitem{Yang2008SR}
J.~Yang, J.~N. Wright, T.~S. Huang, and Y.~Ma, ``Image super-resolution as
  sparse representation of raw image patches,'' {\em 2008 IEEE Conference on
  Computer Vision and Pattern Recognition}, pp.~1--8, 2008.

\bibitem{Bevilacqua2012SR}
M.~Bevilacqua, A.~Roumy, C.~Guillemot, and M.-L. Alberi-Morel, ``Low-complexity
  single-image super-resolution based on nonnegative neighbor embedding,'' in
  {\em BMVC}, 2012.

\bibitem{Chang2004SR}
H.~Chang, D.-Y. Yeung, and Y.~Xiong, ``Super-resolution through neighbor
  embedding,'' {\em Proceedings of the 2004 IEEE Computer Society Conference on
  Computer Vision and Pattern Recognition, 2004. CVPR 2004.}, vol.~1, pp.~I--I,
  2004.

\bibitem{Umehara2017SRmammo}
T.~I. Kensuke~Umehara, Junko~Ota, ``Super-resolution imaging of mammograms
  based on the super-resolution convolutional neural network,'' {\em Open
  Journal of Medical Imaging}, vol.~7, pp.~180--195, 2017.

\bibitem{Umehara2017SRChest}
K.~Umehara, J.~Ota, and T.~Ishida, ``Application of super-resolution
  convolutional neural network for enhancing image resolution in chest ct,''
  {\em J Digit Imaging}, pp.~180--195, 2017.

\bibitem{Plenge2012SRMRI}
E.~Plenge, D.~H.~J. Poot, M.~Bernsen, G.~Kotek, G.~Houston, P.~Wielopolski,
  L.~van~der Weerd, W.~J. Niessen, and E.~Meijering, ``Super-resolution methods
  in {MRI}: Can they improve the trade-off between resolution, signal-to-noise
  ratio, and acquisition time?,'' {\em Magn Reson Med}, vol.~68,
  pp.~1983--1993, 2012.

\bibitem{Ledig2016SRGAN}
C.~Ledig, L.~Theis, F.~Huszar, J.~Caballero, A.~Cunningham, A.~Acosta, A.~P.
  Aitken, A.~Tejani, J.~Totz, Z.~Wang, and W.~Shi, ``Photo-realistic single
  image super-resolution using a generative adversarial network,'' in {\em
  {CVPR}}, pp.~105--114, {IEEE} Computer Society, 2017.

\bibitem{Sanchez2018SRGANmri}
I.~S{\'a}nchez and V.~Vilaplana, ``Brain {MRI} super-resolution using
  generative adversarial networks,'' in {\em International Conference on
  Medical Imaging with Deep Learning}, (Amsterdam, The Netherlands), 2018.

\bibitem{Chuquicusma2017DCGAN}
M.~J.~M. Chuquicusma, S.~Hussein, J.~R. Burt, and U.~Bagci, ``How to fool
  radiologists with generative adversarial networks? a visual turing test for
  lung cancer diagnosis,'' {\em 2018 IEEE 15th International Symposium on
  Biomedical Imaging (ISBI 2018)}, pp.~240--244, 2018.

\bibitem{Adar2018DCGAN}
M.~Frid-Adar, I.~Diamant, E.~Klang, M.~Amitai, J.~Goldberger, and H.~Greenspan,
  ``{GAN}-based synthetic medical image augmentation for increased {CNN}
  performance in liver lesion classification,'' {\em Neurocomputing}, vol.~321,
  pp.~321--331, 2018.

\bibitem{Bermudez2018DCGAN}
C.~Bermudez, A.~J. Plassard, L.~T. Davis, A.~T. Newton, S.~M. Resnick, and
  B.~A. Landman, ``Learning implicit brain {MRI} manifolds with deep
  learning,'' {\em Proceedings of SPIE--the International Society for Optical
  Engineering}, vol.~10574, 2018.

\bibitem{Madani2018DCGAN}
A.~Madani, M.~Moradi, A.~Karargyris, and T.~Syeda-Mahmood, ``Chest x-ray
  generation and data augmentation for cardiovascular abnormality
  classification,'' 2018.

\bibitem{Korkinof2018PGGAN}
D.~Korkinof, T.~Rijken, M.~O'Neill, J.~Yearsley, H.~Harvey, and B.~Glocker,
  ``High-resolution mammogram synthesis using progressive generative
  adversarial networks,'' 2018.
\newblock arXiv:1807.03401.

\bibitem{Wolterink2017CycleGAN}
J.~M. Wolterink, A.~M. Dinkla, M.~H.~F. Savenije, P.~R. Seevinck, C.~A.~T.
  van~den Berg, and I.~I{\v{s}}gum, ``Deep {MR} to {CT} synthesis using
  unpaired data,'' in {\em Simulation and Synthesis in Medical Imaging} (S.~A.
  Tsaftaris, A.~Gooya, A.~F. Frangi, and J.~L. Prince, eds.), (Cham),
  pp.~14--23, Springer International Publishing, 2017.

\bibitem{Hiasa2018CycleGAN}
Y.~Hiasa, Y.~Otake, M.~Takao, T.~Matsuoka, K.~Takashima, A.~Carass, J.~Prince,
  N.~Sugano, and Y.~Sato, ``Cross-modality image synthesis from unpaired data
  using {cycleGAN}: Effects of gradient consistency loss and training data
  size,'' in {\em Simulation and Synthesis in Medical Imaging - Third
  International Workshop, SASHIMI 2018, Held in Conjunction with MICCAI 2018,
  Proceedings} (O.~Goksel, I.~Oguz, A.~Gooya, and N.~Burgos, eds.), Lecture
  Notes in Computer Science (including subseries Lecture Notes in Artificial
  Intelligence and Lecture Notes in Bioinformatics), pp.~31--41, Springer
  Verlag, 1 2018.

\bibitem{Zhang2018CycleGAN}
Z.~Zhang, L.~Yang, and Y.~Zheng, ``Translating and segmenting multimodal
  medical volumes with cycle- and shape-consistency generative adversarial
  network,'' {\em 2018 IEEE/CVF Conference on Computer Vision and Pattern
  Recognition}, pp.~9242--9251, 2018.

\bibitem{Wu2018GANaug}
E.~Wu, K.~Wu, D.~Cox, and W.~Lotter, ``Conditional infilling {GANs} for data
  augmentation in mammogram classification,'' in {\em RAMBO+BIA+TIA@MICCAI},
  vol.~11040 of {\em Lecture Notes in Computer Science}, pp.~98--106, Springer,
  2018.

\bibitem{Mok2019GANaug}
T.~C.~W. Mok and A.~C.~S. Chung, ``Learning data augmentation for brain tumor
  segmentation with coarse-to-fine generative adversarial networks,'' in {\em
  Brainlesion: Glioma, Multiple Sclerosis, Stroke and Traumatic Brain Injuries}
  (A.~Crimi, S.~Bakas, H.~Kuijf, F.~Keyvan, M.~Reyes, and T.~van Walsum, eds.),
  (Cham), pp.~70--80, Springer International Publishing, 2019.

\bibitem{FridAdar2018GANaug}
M.~Frid-Adar, E.~Klang, M.~Amitai, J.~Goldberger, and H.~Greenspan, ``Synthetic
  data augmentation using {GAN} for improved liver lesion classification,''
  {\em 2018 IEEE 15th International Symposium on Biomedical Imaging (ISBI
  2018)}, pp.~289--293, 2018.

\bibitem{Nie2016EDN}
D.~Nie, X.~Cao, Y.~Gao, L.~Wang, and D.~Shen, ``Estimating {CT} image from
  {MRI} data using {3D} fully convolutional networks,'' in {\em Deep Learning
  and Data Labeling for Medical Applications} (G.~Carneiro, D.~Mateus,
  L.~Peter, A.~Bradley, J.~M. R.~S. Tavares, V.~Belagiannis, J.~P. Papa, J.~C.
  Nascimento, M.~Loog, Z.~Lu, J.~S. Cardoso, and J.~Cornebise, eds.), (Cham),
  pp.~170--178, Springer International Publishing, 2016.

\bibitem{Han2017EDN}
X.~{Han}, ``{MR}-based synthetic {CT} generation using a deep convolutional
  neural network method,'' {\em Medical Physics}, vol.~44, pp.~1408--1419, Apr.
  2017.

\bibitem{Kida2017EDN}
S.~Kida, T.~Nakamoto, M.~Nakano, K.~Nawa, A.~Haga, J.~Kotoku, H.~Yamashita, and
  K.~Nakagawa, ``{Cone Beam Computed Tomography Image Quality Improvement Using
  a Deep Convolutional Neural Network},'' {\em Cureus}, vol.~10, p.~e2548, Apr.
  2018.

\bibitem{Avi2019pix2pix}
A.~Ben-Cohen, E.~Klang, S.~P. Raskin, S.~Soffer, S.~Ben-Haim, E.~Konen, M.~M.
  Amitai, and H.~Greenspan, ``Cross-modality synthesis from {CT} to {PET} using
  {FCN} and {GAN} networks for improved automated lesion detection,'' {\em
  Engineering Applications of Artificial Intelligence}, vol.~78, pp.~186 --
  194, 2019.

\bibitem{Kida2019CycleGAN}
S.~Kida, S.~Kaji, K.~Nawa, T.~Imae, T.~Nakamoto, S.~Ozaki, T.~Ohta, Y.~Nozawa,
  and K.~Nakagawa, ``Cone-beam {CT} to planning {CT} synthesis using generative
  adversarial networks,'' 2019.
\newblock arXiv:1901.05773.

\bibitem{Chang_2017_ICCV}
J.~H. Rick~Chang, C.-L. Li, B.~Poczos, B.~V.~K. Vijaya~Kumar, and A.~C.
  Sankaranarayanan, ``One network to solve them all -- solving linear inverse
  problems using deep projection models,'' in {\em The IEEE International
  Conference on Computer Vision (ICCV)}, Oct 2017.

\bibitem{deepimageprior}
D.~Ulyanov, A.~Vedaldi, and V.~S. Lempitsky, ``Deep image prior,'' in {\em
  {CVPR}}, pp.~9446--9454, {IEEE} Computer Society, 2018.

\bibitem{Adler_2017}
J.~Adler and O.~\"Oktem, ``Solving ill-posed inverse problems using iterative
  deep neural networks,'' {\em Inverse Problems}, vol.~33, p.~124007, nov 2017.

\bibitem{Lucas2018}
A.~{Lucas}, M.~{Iliadis}, R.~{Molina}, and A.~K. {Katsaggelos}, ``Using deep
  neural networks for inverse problems in imaging: Beyond analytical methods,''
  {\em IEEE Signal Processing Magazine}, vol.~35, pp.~20--36, Jan 2018.

\bibitem{chainer}
S.~Tokui, K.~Oono, S.~Hido, and J.~Clayton, ``Chainer: a next-generation open
  source framework for deep learning,'' in {\em Proceedings of Workshop on
  Machine Learning Systems (LearningSys) in The Twenty-ninth Annual Conference
  on Neural Information Processing Systems (NIPS)}, 2015.

\bibitem{CPTAC-SAR}
{National Cancer Institute Clinical Proteomic Tumor Analysis Consortium
  (CPTAC)}, ``Radiology data from the clinical proteomic tumor analysis
  consortium sarcomas {[CPTAC-SAR]} collection [data set].''
  \url{https://wiki.cancerimagingarchive.net/display/Public/CPTAC-SAR}, 2018.

\bibitem{Tanno17}
R.~Tanno, D.~E. Worrall, A.~Ghosh, E.~Kaden, S.~N. Sotiropoulos, A.~Criminisi,
  and D.~C. Alexander, ``Bayesian image quality transfer with {CNNs}: Exploring
  uncertainty in {dMRI} super-resolution,'' in {\em Medical Image Computing and
  Computer Assisted Intervention - {MICCAI} 2017 - 20th International
  Conference, Quebec City, QC, Canada, September 11-13, 2017, Proceedings, Part
  {I}}, pp.~611--619, 2017.

\bibitem{ODL}
J.~Adler, H.~Kohr, and O.~\"Oktem, ``Operator discretization library ({ODL}),''
  Jan. 2017.

\end{thebibliography}

\end{document}